# Program structure

**Alex Shkotin**

Cybernetic Intelligence GmbH, Sumfstrasse 26, 6031 Zug,
Switzerland
e-mail: ashkotin@acm.org

**Abstract**   A program is usually represented as a word chain. It is exactly a word chain that appears as the lexical analyzer output and is parsed. The work shows that a program can be syntactically represented as an oriented word tree, that is a syntactic program tree, program words being located both in tree nodes and on tree arrows. The basic property of a tree is that arrows starting from each node are marked by different words (including an empty word). Semantics can then be directly specified on such tree using either requirements or additional links, and adding instructions to some tree nodes enables program execution specification.

**Keywords**   program syntax, program semantics, finite labeled graph

Section 1 contains a summary of the approach and an example of program text and its syntactic tree. The syntactic schema described in Section 2 is used to specify a family of syntactic trees. As a result, the language proves to be a family of trees specified by the schema. Extra requirements depending on programming language semantics are set forth for the syntactic tree of a program. In addition, extra arrows, that is semantic links, are drawn on the syntactic tree of the program. Semantics depends on the programming language, and it will be reviewed in terms of the Turingol language [1] in Section 3. The structure of the external data existing irrespective of the program requires a separate consideration. Section 4 describes data for Turingol alone with program using a tape. A program presented in the form of a syntactic tree with semantic links has to be initialized. For example, the external data to be used by the program has to be connected to it. In addition, instructions are entered for the Executor in nodes corresponding to executable statements. Initialization is described in Section 5, and execution in Section 6.

Appendix 1 describes bringing author's grammar Turingol to a form giving a schema, and Appendix 2 describes tools required to work with finite labeled graphs.

**1 Introduction**

We will review Turingol [1] as an example throughout the paper. It is a simple programming language with formal semantics described since its birth, and notably by the author. However, a closer examination shows that translation of a Turingol program into Turing machine programs, that is a translation verifying the Turingol program, is described. As regards the language itself, the author says that it is clear as it is [1], p.138, lines 1-3. Thus, the requirements for the Turingol program are hidden in translation. However, a couple of requirements are mentioned in express form [1], p.139:

"...programs are malformed if the same identifier is used twice as a label or if a **go to** statement specifies an identifier which is not a statement label."

We will not review Turing machine programs, and will focus on Turingol programs themselves. Some Executor is supposed to execute a Turingol program using a tape. The tape structure will be accordingly described.

A program is a graph of words, and description of its properties and rules of use is the main objective of the paper. An oriented tree of language words, that is a syntactic tree of the program (or its part), underlies the graph. A syntactic schema similar to Wirth [2] syntactic diagrams in a sense is introduced to specify a family of trees. However, while syntactic diagrams specify rules for building a word chain, a schema specifies rules for building word labeled trees. There exist only two ways of creating sub-trees: a sequence of nodes (connected by arrows) and a node with a set of arrows specified by the schema starting from it. The fact that a particular schema generates trees with differently labeled outgoing arrows has to be proved, which will be done for Turingol.

The fact that the program itself is a tree enables a new view of the description of its semantics. Various attributes may obviously be assigned to tree nodes, including assignment for compilation [1].

The obtained tree has to meet some requirements providing a well-formed Turingol program [1], p.138. We will assume for simplicity that there exists a separate verification phase for these requirements, although any verifications can undoubtedly be carried out in the course of building up a tree.

If the tree is good, then additional arrows expressing syntactic and semantic relationships facilitating Executor's work will be drawn between some of its nodes. Tree completion (to a graph) is highlighted as a separate phase for simplicity as well.

To completely understand a program, one has to specify its execution procedure. An abstract Executor moving across the program graph and executing instructions located in nodes corresponding to statements is supposed to exist for this purpose similar to [6]. This approach corresponds to how a programmer thinks of his program. Placing instructions in nodes is included in the program initialization phase. It also includes program connection to external data, which consists of connecting a tape to the program in case of Turingol.

The Executor executes the program by moving from some of its nodes to other and executing the instructions specified there (including operations on the tape). This is exactly what the programmer had in mind when creating the program. The idea is that the programmer can basically execute his program on his own without using a machine (for example, Turing machine). Moreover, it is exactly this understanding of how an Executor executes the source text of the program on the part of the programmer that obliges such Executor (aka debugger) to interact with the programmer as if it (Executor) were executing the source text of the program.

**1.1 Program example – Text**

Let's review program 4.1 [1], p.137. Some word combinations are written through a dash for simplicity, which makes them lexically one composite word. The dash has to be accordingly added to the language alphabet.

tape-alphabet is blank, one, zero, point;
print "point";
go to carry;
test: if the-tape-symbol is "one" then
{print "zero"; carry: move left one-square; go to test};
print "one";
realign: move right one square;
if the-tape-symbol is "zero" then go to realign.

**1.2 Syntactic tree of a program**

Syntactic tree of program 4.1 is given in Fig. 1.

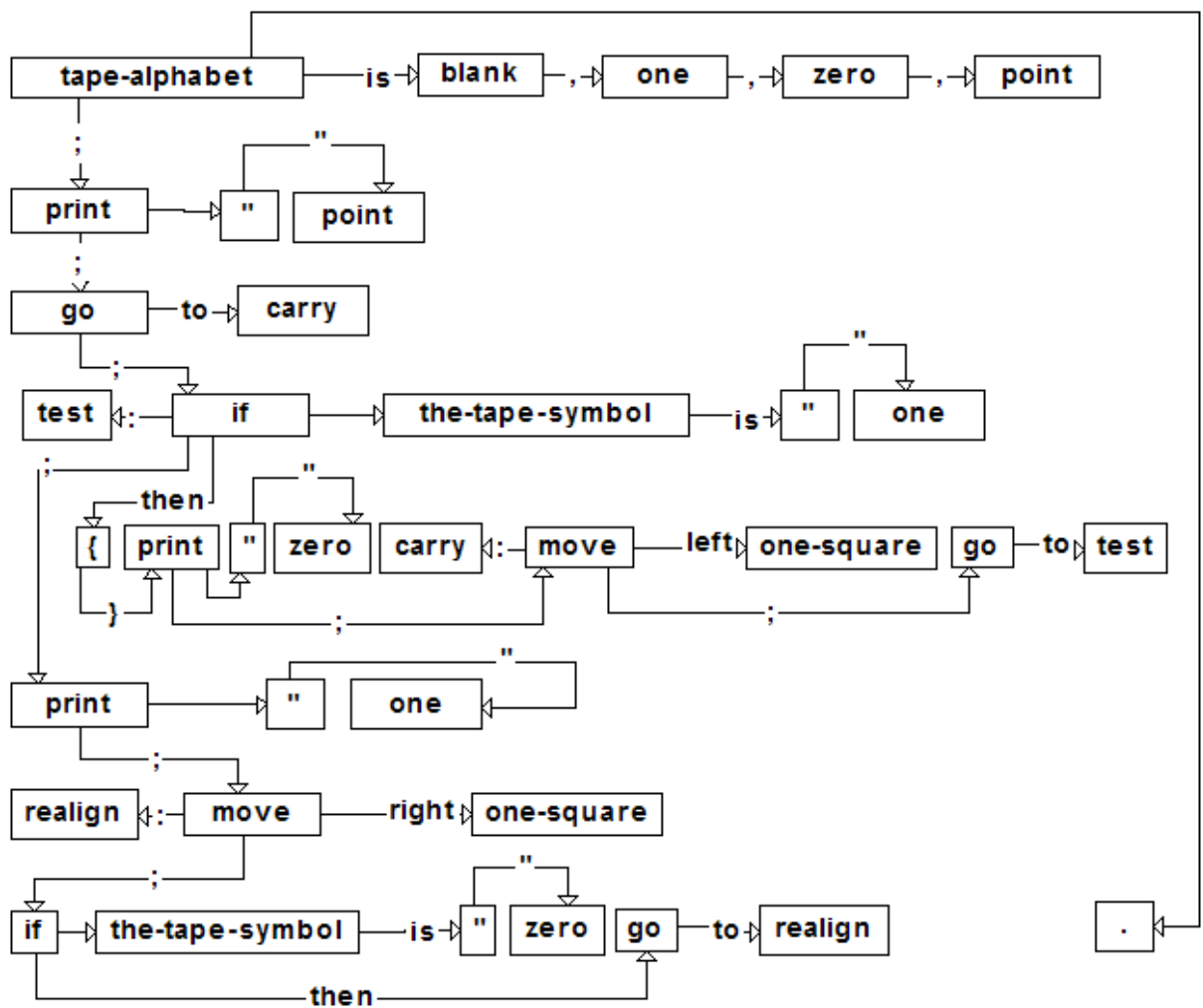

**Fig. 1** Syntactic tree example

Each program word (including special words such as ';', ',') goes to a specific label (of a node or an arrow). The arrow label and generally node label as well may be an empty word.
One may see for oneself that arrows starting from any node are differently labeled.
Let's highlight node chains by a separator placed on arrows:

First row: 'blank' ',' 'one' ',' 'zero' ',' 'point'.
Fifth row: 'print' ';' 'move' ';' 'go'.
Vertically: 'print' ';' 'go' ';' 'if ' ';' 'print' ';' 'move' ';' 'if'.
The first chain is uniform in that its members are of identical structure. This is simply a word in this case. The second and third chains are non-uniform, as their component statements have differently structured trees.

Note, too, that the program tree is not ordered, i.e. it contains no child node order, an order that could be used to specify tree drawing (position on a plane or line).

We will see that such tree contains enough information for the entire program semantics. Accordingly, one may assume that drawing rules are only used for convenience of recognizing parts of the tree.

## 2 Schema. Specifying tree families

A syntactic schema is used to specify families of labeled trees.
Regular expressions are used to specify permissible values of a word in a node or on an arrow of a tree. However, a specific word only is usually permissible. We will need only two regular expressions: left|right and [a-z]+ for Turingol, the last one specifying a word over an alphabet of lower-case English letters.

An explanatory comment will be further added in square brackets to the text of definitions, which will not be a part of the definition formally.
Let PLA and MLA be two non-overlapping alphabets.
PLA is a programming language alphabet, and MLA is a metalanguage alphabet.

**Definitions**
*Syntactic tree* (*sytr* in abbreviated form) is oriented labeled tree such that the node label together with arrows are a word from PLA*.
*Uni-labeled tree* is sytr with arrows starting from each node labeled differently.

*Sentential tree* (setr) is oriented labeled tree such that:
  1. Node label is a word from PLA* or MLA+
  2. Arrow label is a word from PLA*.
Nodes labeled by a word from MLA+ are named *auxiliary* and play a role similar to non-terminals in context-free grammars.

*Syntactic schema* is oriented labeled graph with 2 types of arrow (*AND arrows* and *OR arrows*), AND arrows being divided in two (*mandatory/optional*) subspecies.
Labels are regular expressions over PLA [including words from PLA*].
A node with no outgoing arrows is named an *atomic node / A node*.
A node with outgoing OR arrows only is named a *OR node*.
A node with outgoing AND arrows only is named a *AND node*.
Each schema node can be assigned a unique name (*schema name*), which is a word from MLA+. Such schema is named a *completely named schema*.
**End of definitions**

## 2.1 Turingol. Schema

Let us review a completely named Turingol schema as an example (see Fig. 2).

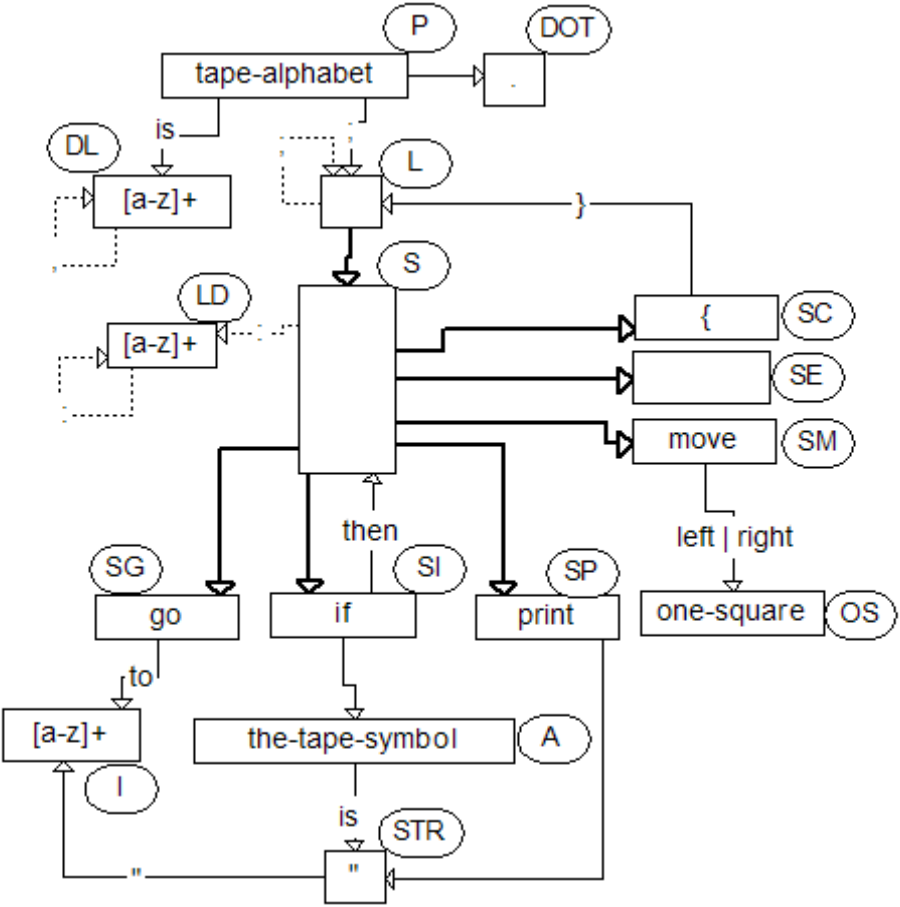

**Fig. 2** Syntactic schema for Turingol

Nodes are rectangles. Ovals contain schema names. OR arrows are marked by heavy lines. Optional arrows are marked by dash lines. The majority of nodes and arrows are labeled by specific words. DL, LD, I nodes are labeled by the regular expression [a-z]+. An arrow from SM to OS is labeled by the regular expression left|right. L, S, SE nodes are labeled by empty words. P-DOT, SI-A, SP-STR arrows are labeled by empty words too.

Let us now describe how a syntactic schema specifies a sytr family.

*Building a sentential tree from a schema node*: Let the schema be somehow completely named, and a schema Y0 node specified. To create a sentential tree from the Y0 node,
Create an isolated N1 node (a Y0 node copy).
Draw copies of all mandatory and some optional Y0 node outgoing AND arrows from the N1 node, and label arrow copies by words permissible by regular expressions of their samples in the schema. Create a node with a schema name of the corresponding schema node at the end of each arrow. If an Y0 node has OR arrows, then take a node name at the end of one of them, and enter it in N1; otherwise, label N1 by a word specified in the regular expression of the Y0 node.

The instruction becomes significantly simpler for an OR node:
Create an isolated N1 node.
Take a node name at the end of a Y0 node outgoing OR arrow and enter it in N1.
The instruction for an atomic node is simple as well:
Create an isolated N1 node.
Label N1 with a word specified in the regular expression of the Y0 node.

Thus, the schema enables assignment of a set of sentential trees to each schema name of a node. We actually have a context-free grammar of trees in a compact form where the left part of the rule contains an isolated node labeled by a schema name, and the right part, a setr or sytr.
For example, two setr are generated from L (because of an optional arrow): one is simply S, and the other is S with a ';' arrow in L; and 12 setr are generated from S: six isolated nodes (SG, SI...) and six such nodes with a ':' arrow in LD.
OR arrows (including one as in L) enable specification of setr with a non-terminal in the root.
A tree is inserted in another tree by substituting the root of the other tree for a node of the first one.
The schema is convenient due to the fact that it is a connected graph, and so enables an association between some graph properties and properties of the family of generated trees.

*Building a sytr using a schema*: Let a schema be completely named. To obtain an initial setr, it suffices to take any schema node, and to create an isolated node labeled by the schema name of this schema node.
Let there be given a C1 setr with a N1 node labeled by any schema name. Then a C2 setr can be created from a schema node with a schema name equal to N1 label and substituted for N1 in C1.
If no node labeled by a word from MLA+ is found after several substitutions in C1, then we have obtained a sytr.

*Schema properties*: one can readily see that:
1. No OR arrow label is used
2. Parallel OR arrows are redundant
3. The label of a node with an outgoing OR arrow is not used.

Thus, we can assume without loss of generality that:
1. The label of an OR arrow or a node with an outgoing OR arrow is an empty word
2. There are no parallel OR arrows on the schema.

Single AND loops (i.e. when a node has a single loop) play an important role, as they specify chains. Parallel AND loops specify a tree, i.e., a structure that apparently does not occur in programming languages. Note that if an AND arrow of a loop is mandatory, then the building process will never come to an end, and this part of the schema is useless for finite trees. Thus, all AND loops can be considered optional.

## 2.2 Uni-labeled families

So, a schema specifies a family of trees. We can try to prove that only uni-labeled family trees exist by analyzing the schema. The following schema property is required for a family to be uni-labeled.

*AND condition*: regular sets of regular expressions of AND arrows starting from each node do not overlap.

The AND condition can be easily checked on a schema. But it is insufficient, as if both AND arrows and OR arrows start from a N1 node, then AND arrows from nodes at OR arrow ends, etc. may be added to N1 AND arrows. For example, a ':' AND arrow from S (*statement label*) can propagate onto any statement by following any OR arrow in case of a Turingol schema.

The most important property of 'interesting' schemas is presence of cycles on the graph. The cycle can be characterized by the node types it includes.

OR loop possesses the following property: if it is located in an OR node, then it is redundant, otherwise it results in a non-uni-labeled sytr, as AND arrows will be drawn from the node in the course of building, and the same 'non-terminal' will remain in the node, which means that the same arrows can be drawn from it again. An OR cycle (OR arrow cycle on a schema graph) gives non-uni-labeled trees as well. Therefore, we will be interested in schemas with *AND cycle condition* satisfied: each cycle includes an AND node. It also means that the schema has no OR loops.

If the AND cycle condition is satisfied, a simple algorithm of AND arrow label propagation on the schema works: we create a <node schema name, arrow label> pair for each AND arrow, and place it in the node located at the origin of the AND arrow. Each pair so created follows all OR arrows and 'settles' on AND and A nodes.

*Sufficient condition*: If all of the pairs accumulated in AND and A nodes (including those initially existing) after 'advance' have pairwise disjoint regular sets (no conflict), then the schema generates only uni-labeled trees.

The Turingol schema satisfies the AND condition and the AND cycle condition, as there are only two cycles with OR arrows on the schema – SC-L-S-SC and SI-S-SI, each containing an AND node (SC, SI, respectively).
The propagation algorithm results in <L,';'> and <S,':'> pairs arriving in nodes at the ends of OR arrows from S, but without conflicts.
Thus, the sufficient condition is satisfied, and the schema specifies a family of uni-labeled trees.

Clearly, an arbitrary schema may generate exotic trees, including infinite ones. Appendix 1 discusses the relationship between the schema and context-free grammars, including demonstration of the Turingol context-free grammar required to obtain a schema.

### 2.3 General case of a programming language

Following [5] and exercise 2.4.28, "2.4 Context-Free Languages" from [4] each context-free language is produced by a grammar with all rules looking like A:aBbC, A:aBb, A:aB, A:a. If the empty word belongs to the language, then the S:e rule is permitted, where capital letters denote some non-terminals, lower case letters (except 'e') – some terminals, and 'e' – the empty word.
It can be easily seen that a sentential tree can be assigned to each type of the right part. If we present the graph as sets of triples <arrow origin, arrow label, arrow end> corresponding to the number of arrows in graph, then we obtain:
- two triples: <a " B>, <a b C>., i.e. 'a' is a root with an empty word labeled arrow to B and a 'b' labeled arrow to C for the first type of rule.
- one triple: <a b- B>. for the second type of rule. Where '-' after b means here that b

should follow B linearization during tree linearization. Without a sign "-" a situation is just reverse.
- one triple: <a " B>. for the third type of rule.
- The fourth and fifth rules give atomic nodes.

The context-free grammar form given above is named a "standard operator form" [5]. Thus, if we present a grammar as a standard operator form, then we can mechanically go over to uni-labeled trees. Every programming language can be theoretically said to permit a representation of its programs as uni-labeled trees. There clearly exist several such representations, and some of them are likely to be visual and natural. The Author of the language had best to take care of a representation in the form of trees.

## 3 Turingol. Program structure

There exist requirements already for the syntactic tree of the program.
In addition, the following objects converting a sytr to a graph have to be additionally built on the syntactic tree of a program:
- Semantic link: 'is-declared-at' arrows
- Semantic control flow links: 'next', 'yes', 'no' arrows.

**Designations**: the requirements are named. The requirements that are not critical for program execution have a W (warning) letter in their names.
Appendix 2 contains some ways of handling the graphs that we will need below.

### 3.1 alphabet

**Definition**: tape word declaration node (*w-declaration-point*) is node in a chain with a **'tape-alphabet'+is** head.
**'tape-alphabet'+is** is a path formula (see Appendix 2) and means a node at the end of an 'is' labeled arrow starting from a 'tape-alphabet' labeled node, i.e. one is proposed to follow (indicated by '+') the 'is' labeled arrow from a 'tape-alphabet' labeled node.
**Definition**: tape word use node (*w-usage-point*) is any **print**+'"' node or **if**+''+**is**+'"' node, **print**+'"' and **if**+''+**is**+'"' being path formulas here (see Appendix 2).
The fact that these paths are meaningful (passable) can be easily seen from the schema.

**AW1, AW2, AW3 requirements**

(AW1) Labels must be different in all w-declaration-points.
(AW2) Label value must be equal to the label of a w-declaration-point in any w-usage-point.
(AW3) The label of each w-declaration-point must be used in a w-usage-point.
Note: It is not so in Example 4.1, and it may be a cause for warning.

### 3.2 'is-declared-at' arrow for a word usage point

Once fulfilled, the requirement (AW2) enables drawing an 'is-declared-at' arrow from a w-usage-point to a w-declaration-point.
Note: The Executor does not need such arrows.
They are required for further processing in other languages.

## 3.3 Label

**Definition**: Label target point (l-target-point) is any ':' arrow destination node.
**Definition**: Label use place (l-usage-point) is any 'to' arrow destination node.

### L1, L2, LW1 requirements

(L1) All labels must be different in l-target-points.
(L2) Every l-usage-point must have an l-target-point with the same label.
(LW1) Every l-target-point must have an l-usage-point.
Otherwise, it is useless.

## 3.4 Control flow graph

Let there be given a sytr.
The word 'go' may prove to be also a label or a tape word. To distinguish, let us introduce the simplest classification on sytr:
*Data node*: this is a node of a sub-tree starting from a **'tape-alphabet'+is** node as well as the node at the end of the ':', 'to', '' arrows.
*S node*: this a non-data node labeled by the word 'go', 'if', 'print', 'move', '', '{'.
*L-chain member*: a ';' arrow destination member.
Let's also introduce a simple S node classification: *control statements* are 'if', '{`, 'go' labeled nodes. Let us name the other *ordinary* ones: 'print', 'move', '' labeled nodes.

Once the Executor has executed an instruction in the next in turn S node, it must know where to go further, and so forth until it has executed a Stop direction. The next executable node is one and only one for the majority of the nodes, across which the Executor moves. We will draw a semantic 'next' arrow to the next executable node. Only 'if' nodes are exceptions in that two control arrows 'yes', 'no' will be drawn from them similar to the situation described by Post [6]: "(B) Perform operation (e) and according as the answer is yes or no correspondingly follow direction $j_i$ " or $j_i$ ",".

The Executor starts from the 'alphabet-tape' node, and must syntactically follow the ';' arrow to the first statement. It is a frequent situation when the ('next', 'yes', 'no') control arrow is parallel to the syntactic arrow.

*Building procedure:* Draw a 'next' arrow in parallel to a ';' arrow starting from the 'alphabet-tape' node.

Program halting is a special act that is only implied in a language such as Turingol, i.e. it has no express statement. Let's introduce a special additional node named *Stop* where the Stop direction will be placed at the program initialization stage (see below) to simplify program structure and to follow Post's ideas [6].

*Building procedure:* Create a node and label it by the word 'stop'.

The 'tape-alphabet' and '{' nodes have a subordinate statement chain; 'if' node has one subordinate statement. The next statement to be executed after an ordinary statement located last in an L-chain or subordinate to 'if' is specified in the subordinating node, provided that the subordinating node is not a subordinate one, otherwise one must go to its subordinating node.

Let's draw an auxiliary 'back' arrow to show the subordination relationship.

*Building 'back' arrows*: Draw a 'back' arrow from the last member to the node subordinating the chain (i.e. 'tape-alphabet' or '{' node) in each L-chain. Draw a 'back' arrow from each node hanged to 'if' node over 'then' arrow to this 'if' node.

Thus, it follows from the building procedure that one and only one of the situations listed below occurs for each S node:
*BACK situation*: there is a 'back' arrow starting from it
*';' situation*: there is a ';' arrow starting from it, i.e. the statement is not the last one in the L-chain.

*Building procedure:* Redirect the 'back' arrow with an 'alphabet-tape' destination node to the 'stop' node.
The result of building 'back' arrows for Example 4.1 is presented on the Figure 3.

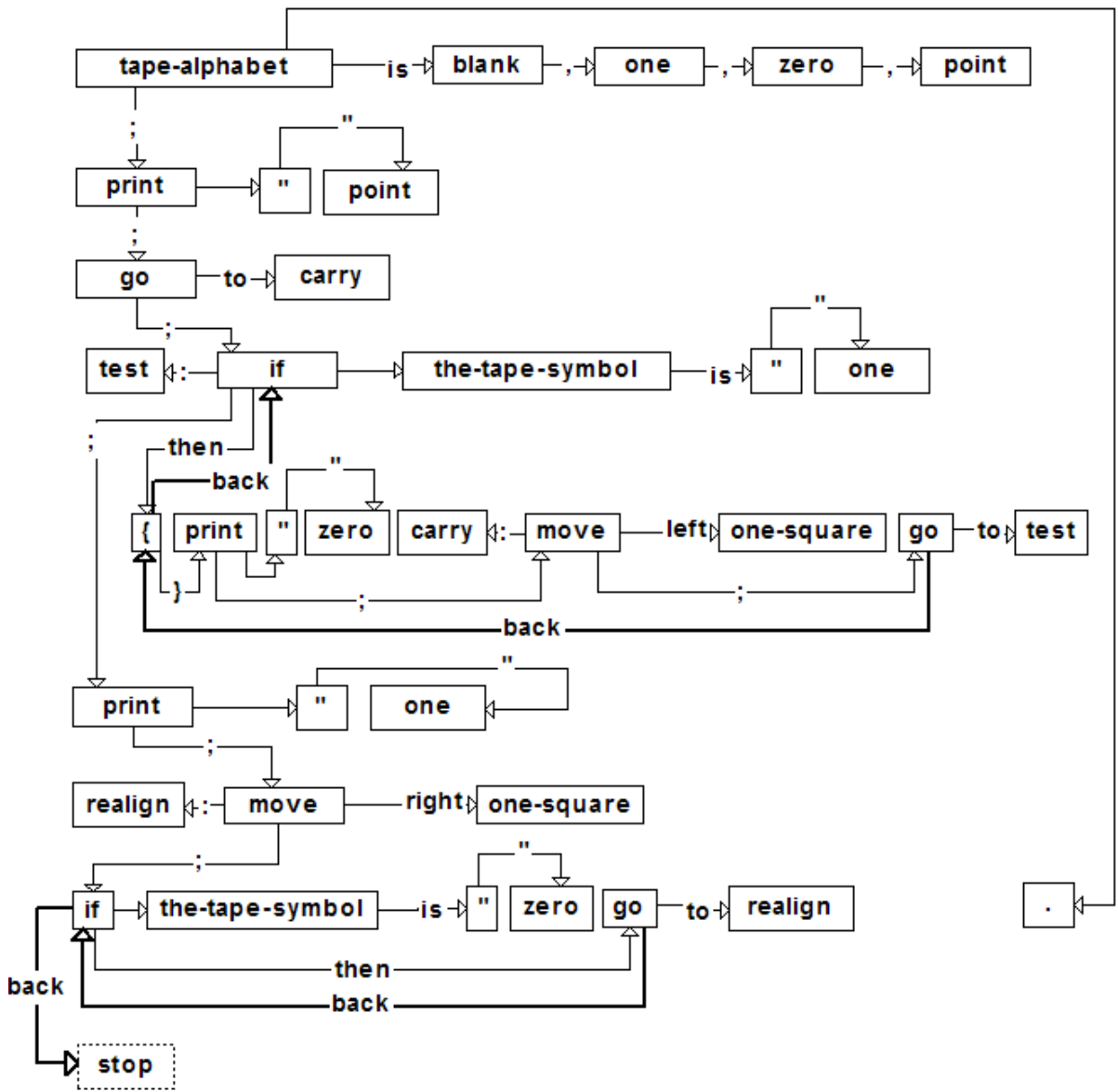

**Fig. 3** Syntactic tree with 'back' arrows

*Building control arrows that do not depend on the BACK or ';' situation*:
Draw a 'yes' arrow in parallel to 'then' arrow for each 'if' node.
Draw a 'next' arrow in parallel to '}' for each '{' node.
For each S node labeled 'go', using its l-usage-point, draw a 'next' arrow to the last node, to which one can rise from an l-target-point with the same label value as the l-usage-point label along ':' arrows.

The possibility of the last building procedure is ensured by fulfillment of the L1, L2 requirements.

*Building control arrows for the ';' situation*, i.e. a statement that is not last in the L-chain:
Draw a 'next' arrow parallel to the ';' arrow for ordinary statements ('', 'print', 'move').
Draw a 'no' arrow parallel to the ';' arrow for the *if* statement.

Thus, the syntactic ';' arrow is not used to build 'next' arrows for 'go' and '{' nodes in the ';' situation.

*Building the control for nodes with an outgoing 'back' arrow*:
1. Label all 'back' arrows as unprocessed ones
2. Stop if there are no unprocessed 'back' arrows
3. Select the entire (full length) unprocessed 'back' arrows chain. It either goes to ';' (let's name it a *NEXT situation* and this ';' arrow a *C1*) or ends in the 'stop' node (let's name it *a STOP situation*)
Draw a 'next' arrow from each ordinary node of a 'back' arrows chain with an outgoing 'back' arrow, and draw a 'no' arrow from the 'if' node. Draw this arrow:
- To the same node that is the C1 end node in the NEXT situation
- To the 'stop' node in the STOP situation.
Label all 'back' arrows chain arrows as processed ones. Go to 2.

We have obtained a control flow graph. It is exactly the one, along which the Executor moves when executing a program.
The control flow graph is built on S nodes, 'alphabet-tape' node, which is a start node, and 'stop' node. Thereby, it is built on syntactic nodes (plus 'stop' node) and is simply structured from the viewpoint of outgoing arrows:
- One and only one 'next' arrow starts from each S node (except 'if' node) and 'alphabet-tape' node.
- One and only one 'no' arrow and one and only one 'yes' arrow starts from each 'if' node.

No control arrows parallel to syntactic arrows are drawn on the drawing below (see Fig. 4), which corresponds to Example 4.1, but corresponding syntactic arrows are drawn bold. The label of the corresponding control arrow can be easily restored. So, 'no' arrow obviously corresponds to the ';' arrow from 'if' in line 4.

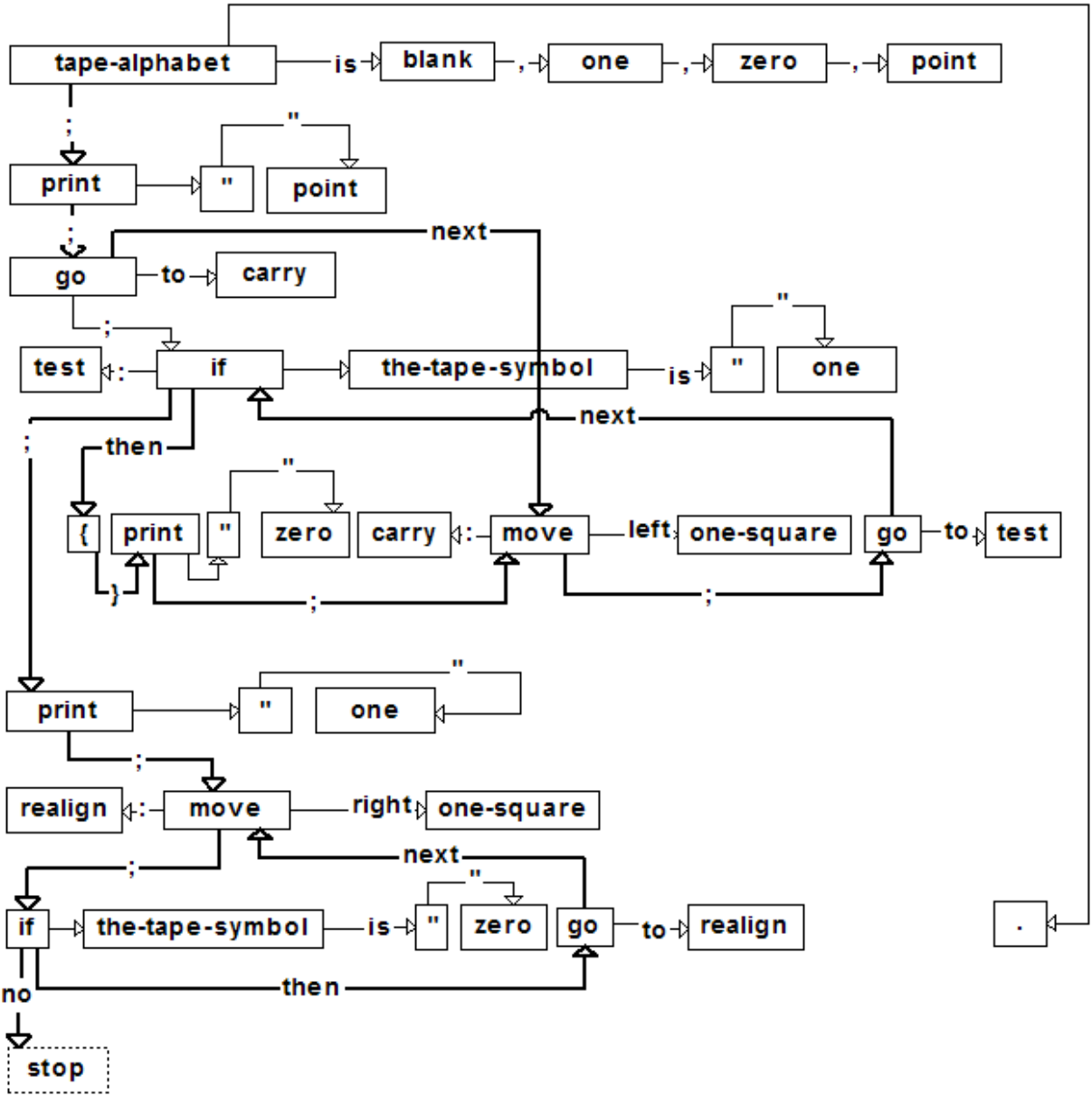

**Fig. 4** Syntactic tree with control flow graph

**CW1, C2 requirements**

The idea of control flow generates a natural requirement:
(CW1) Reachability of any program S node from the 'alphabet-tape' node across the flow graph.

For example: If there is a 'go' node in the L-chain, then the node located at the end of the ';' arrow starting from it must have a label, otherwise it is unreachable. See, for example, a ';' arrow from 'go' node in the third line in 4.1.

The problem of halting a program on data (including any kind of data) is extremely important, as indicates program usability. It has to be solved specifically for each program or a class of programs. However, there are also obvious negative results requiring only an understanding of the idea of control flow.

Requirement (C2): 'next' arrows may not form a cycle.
In particular, "go to" may not point at itself. Note that C2 is no CW1 consequence, as several control arrows may belong to a node.

## 3.5 Conclusion

The presence of a construct in the form of a tree enables a precise and natural expression of requirements for a program as well as setting rules for building additional links.
'is declared at', 'back', 'next', 'yes', 'no' links as well as 'stop' node are necessary in the majority of programming languages.
Whether any links else are necessary is the subject of ongoing study of Standard Pascal [8].
One of the candidates is the 'has type' link for languages with data types.
The flow graph is easier to 'translate' to a Post machine than a Turing machine.

## 4 Data. Tape

Tape is a finite chain of nodes labeled by words. Arrows are labeled by empty words.
Tape finiteness necessitates its completion if a step to the left or to the right is required to a non-existing node. It has to be done as required. The node and the arrow are labeled by empty words at creation.
Note: Such tape is obviously no Turing machine tape, as it:
- Contains words, not letters in cells
- Is finite (see [6] p.105, about possible tape improvements)

A family of tapes, which is a syntactic construct, may be formally described in addition to language grammar:
T:: =I | I T
where I stands for identifier non-terminal (see Appendix 1).
The description of a language actually requires two starting symbols – for the program and for external data.

### 4.1 Operations on the tape

The Executor will:
- Compare labels of some program nodes with labels of tape nodes
- Put labels of some program nodes on tape nodes
- Expand the tape

Tape expansion is based on the following situations and operations.
Let N1 be tape root node name.
Then execution of the "Create a node and an arrow from it to the n1 node." direction (see Appendix 2) will give a tape again with a new root and an arrow from it to the old root.
Let N2 be the last tape node name.
Then execution of the "Create a node and an arrow from the n1 node to it." direction (see Appendix 2) will give a tape again with a new last node.

### 4.2 Conclusion

The tape used by the Turingol program is a prototype of the programming language file, and the method of representation in the form of a chain may be applied to the file as well.

# 5 Initialization

A 'tape' labeled arrow is created from the 'tape-alphabet' node to some node of the tape to be used by the program at the start. The tape node containing the end of 'tape' arrow is named *current tape node*.

The current tape node is usually supposed to be the first tape node 'after tape opening'. However, the last tape node must be the initial current node for the program from Example 4.1, for example.

In addition, instructions for the Executor are entered in some program nodes. Instructions are placed in nodes in a way similar to Post's approach [6] except that programs consist of numbered lines with branches on numbers according to Post, while we assume that programs consist of nodes with branches on arrows.

Note: These instructions may be assigned to schema nodes as attributes and be copied to sytr already at the building stage.

## 5.1 Instructions in nodes

Instructions are given below for S nodes, **'tape-alphabet'** node and 'stop' node. Other nodes have no instructions.

Two instruction versions exist for 'move' nodes.

It is important to stress that the nature of the operations performed in instructions is working with labeled graphs by changing node labels, creating nodes and arrows, and reassigning the destination of the 'tape' arrow. If the terminology is not intuitively clear, then see a more precise description in Appendix 2.

**Tab. 1** node instructions

| node label | instruction |
|---|---|
| tape-alphabet | Follow the 'next' arrow. |
| stop | Stop. |
| go | Follow the 'next' arrow. |
| { | Follow the 'next' arrow. |
| if | If the **'tape-alphabet'+tape** node label equals +"+**is**+"" node label, then follow the 'yes' arrow.<br>Follow the 'no' arrow. |
| " | Follow the 'next' arrow. |
| print | Label the **'tape-alphabet'+tape** node by a +"+"" node label.<br>Follow the 'next' arrow. |
| move | A version of the instruction for a 'move' node with an outgoing 'left' arrow:<br><br>If no " arrow exists to a **'tape-alphabet'+tape** node, then create a node and an arrow from it to the **'tape-alphabet'+tape** node.<br>Reassign the 'tape' arrow to a **'tape-alphabet'+tape-"** node.<br>Follow the 'next' arrow. |

| node label | instruction |
|---|---|
| move | A version of the instruction for a 'move' node with an outgoing 'right' arrow:<br><br>If no '' arrow exists from a **'tape-alphabet'+tape** node, then create a node and an arrow from the **'tape-alphabet'+tape** node to it.<br>Reassign the 'tape' arrow to a **'tape-alphabet'+tape+''** node.<br>Follow the 'next' arrow. |

*Note*: the unique place where a passage against the course of an arrow was necessary and the '-' operation was used is the **'tape-alphabet'+tape-''** path formula, i.e. jump from a 'tape-alphabet' labeled node following a 'tape' arrow to the node located at its end (current tape node), and then jump from the current node against an empty word labeled arrow to the node at its origin. This latter node may be named a *previous node*.

'Code optimization': two complete sets of instructions exist for 'move' nodes. The selection has to be based on the neighboring nodes. In continuation of this subject, one may take the label of a node located at the end of the **+''+is+''''** path from an 'if' node, and place it on an appropriate place in the instruction itself. 'print' nodes may be handled in a similar way.

### 5.2 Properties of instructions

Instructions possess properties, which jointly ensure a fault-free execution of any program on any tape.

Let's list some basic properties.

On the whole,
1. Only instructions located in 'move' nodes modify the graph structure through tape expansion. The tape chain structure is not violated.
2. Only instructions located in 'print' nodes modify graph labeling and affect tape nodes only.

In addition,
1. Each instruction takes the Executor away from a node, i.e. the Executor cannot find itself in a DIRECTIONS EXHAUSTED IN THE NODE situation.
2. Each instruction located in an S node takes the Executor to an S node or 'stop' node. So, the Executor always reaches a node where an instruction is present and cannot find itself in a NO INSTRUCTION IN THE NODE situation.
3. Any path formulas are passable in instructions. Some of them are always passable due to the building procedure, as, for example, the +''+'''' path from a 'print' node or the +''+**is**+'''' path from an 'if' node. The paths that include a 'tape' arrow during program execution point at different tape nodes, but always remain passable at the time of use.

## 6 Execution

The Executor executes a program starting from a node and executes the instructions assigned to the next node. Execution may generally fail. A fail-safety hypothesis exists for Turingol.

Initial state of the program-plus-tape construct:
1. predefined tape node is connected to the 'tape-alphabet' node using the 'tape' arrow.
2. Instructions are entered in nodes.

The Executor starts from the 'tape-alphabet' node.

The Executor moves from node to node in the directions shown by arrows. The Executor executes the instructions contained in the nodes reached.
If a node contains no instruction, then the Executor fails with a No Instructions situation. A similar crash occurs in the Instructions Exhausted in the Node situation.

Each direction has its necessary and sufficient condition of normal execution. The Executor generally must act cautiously and verify the necessary and sufficient condition of normal execution before executing a direction. It will enable it to report a crash.

However, the following *fail-safety hypothesis* exists for Turingol:
Let C1 sytr be built on a schema, meet requirements, and be supplemented with links and instructions.
Let T1 be a tape.
If an arrow is drawn from the 'tape-alphabet' node to any T1 tape node, and the Executor is started in the 'tape-alphabet' node, then the process will never fail.

For example, INSTRUCTIONS EXHAUSTED IN THE NODE SITUATION may not occur, as the last direction in each instruction is FOLLOW THE ARROW form.

Thus, despite the fact that instruction elements (including path formulas) may fail in an arbitrary situation, Turingol computer system (program+data) possess a property that is unusual for programming languages, i.e. fail-safety.

One of interesting tasks is to develop Hoare axioms [7] for operations and to prove the fail-safety hypothesis.

## 7 Discussion

The program structure obtained seems rather realistic, as the Programmer had links such as 'next' and 'is declared at' in mind when creating the program, and exactly the action that the Executor would later find in a node when writing down a statement.

The Executor is a person under the described approach, as instructions are written in a limited natural language. The Executor must be able to work with labeled graphs and certainly letter strings. The computer system structure itself, i.e. a program graph filled with instructions and connected to data, is acceptable so far as we can prove its properties.

It is interesting that a uni-labeled tree is enough for semantics, i.e. we did not need an ordered tree. Some agreements on drawing are most likely required for recognition on the drawing graph elements (as it is always with graphs).

Several sytr versions obviously exist for a programming language. The author of the language had best provided a version of his own.

The fact that the tape proved to be a graph is a start of a study where both external and internal programs data is supposed to be represented by labeled graphs, which are usually trees.

*Optimization and Post machine*: Program graph may certainly be control optimized, i.e. many paths may be reduced. As a result, we will obtain an isomorphic Post machine structure.

*Knuth attributes (compilation)*: Attributes may be specified for nodes already on the schema, and a sytr with attributes enables using the Knuth method for calculation of their values – for example, for compilation.

Although regular expressions are mentioned in schema definition, just a few regular expressions are used in programming languages in practice:
- Specific words of the language (including the empty word) or their finite |-combinations,
- Identifiers,
- Numbers (integers, real numbers).

A finite labeled graph as a mathematical construct enables definitions, algorithms and proofs. All these things may be formalized if required, but that was not the purpose of this work.

### 7.1 Conclusion

The author of the language could have describe the tree structure at once. The author of the language could precisely specify semantic requirements for the program on the tree although additional effort is required to do that. A precise language semantics description must results in high-quality and even intelligent compiling programs.

In studying a phrase of the language, the programmer should probably know not only the order of terms, but also the main one and the one to be put on an arrow. However, a suitable tree editor may hide the tree structure from the programmer.

If a tree editor is created, then building and storing a program in the form of a tree will make the lexical analysis and parsing unnecessary.

One can briefly say that while a program is usually considered to be a word chain, it actually proves to be a uni-labeled word tree.

**Appendix 1: Relationship between a schema and a context-free grammar**

### A1.1 From schema to grammar

A sytr has to be ordered to represent it in the form of a word chain. Node and outgoing arrow numbering is introduced on the schema for that purpose: The node and AND arrows starting from it are numbered by numbers starting from 1. OR arrows remain unnumbered.
If an AND arrow word has to be located after a word chain generated by a node at the end of the arrow, then '-' letter is put after the number.
Algorithm for building a production on the basis of a schema node:
Left part: Node schema name
Right part: List regular expressions on AND arrows and node schema names at their ends in the order of numbers on arrows. If an arrow number contains '-', then the regular expression of the arrow is written after the schema name. If an arrow is optional, then a meta-statement of optionality is assigned to the pair.
The regular expression of the node is put under node number in the event that there are no OR arrows; otherwise, the - |-expression from node schema names on OR arrow ends is put there.

Examples:
Let the numbering for L to be as follows: 1 for the L node and 2 for the ';' arrow.
We obtain:
```
L::= S (';' L)?
```

Let the numbering for S to be as follows: 2 for the S node and 1 for the ':' arrow.
We obtain:
```
S::= LD ':' (SG|SI|SP|SM|SE|SC)
```

### A1.2 Turingol. From grammar to schema

To obtain a syntactic schema, we will start with the author's version of the grammar, convert it to the necessary form, and show how the grammar should be represented in the form of a schema generating syntactic trees of Turingol programs.
EBNF notation from [3] is used to record context-free grammars.
Author's context-free grammar Turingol (see Table 1 in [1]).

```
A::= [a-z]
I::= A | I A
D::= 'tape alphabet is' I | D ',' I
O::= 'left' | 'right'
S::= 'print' '"' I '"' | 'move' O 'one square' | 'go to' I
    | '' /*empty word*/
    | 'if the tape symbol is' '"' I '"' 'then' S | I ':' S
    | '{' L '}'
L::= S | L ';' S
P::= D ';' L '.'
```

Some insignificant changes in the language lexicon:
Composite terms are written through a hyphen (for example, tape-alphabet)
So, we will need an additional letter '-'.
**Alphabet**:

```
PLA:: = [-a-z:;{}.]
```

**Equivalent conversion of the grammar**: We need rules with right parts possessing certain properties – roughly speaking, we need non-terminals to be alternated with terminals.

Transformation:
1. D, O non-terminals are substituted in their places of use.

2. For the L non-terminal equivalent rule is
```
L::= S (';' L)?
```
Let's introduce for P
```
DL::=[a-z]+ (',' DL)?
```
We obtain:
```
I::= [a-z]+

S::= I ':' S | 'print' '"' I '"'
     | 'move' ('left' | 'right') 'one-square'
     | 'go' 'to' I | ''
     | 'if' 'the-tape-symbol' 'is' '"' I '"' 'then' S
     |  '{' L '}'
P::= 'tape-alphabet'  'is' DL  ';'   L '.'
```
3. Let's introduce
```
LD::=[a-z]+ (':' LD)?
```
Then
```
S::= I ':' S | Q
```
can be written as
```
S::= (LD ':')? Q
```
which gives
```
S::= (LD ':')?
     ( 'print' '"' I '"' | 'move' ('left' | 'right') 'one-square'
     | 'go' 'to' I | ''
     | 'if' 'the-tape-symbol' 'is' '"' I '"' 'then' S
     |  '{' L '}'
     )
```

4. Let's introduce additional non-terminals:
```
OS ::= 'one-square'
DOT::= '.'
STR::= '"' I '"'
A  ::= 'the-tape-symbol' 'is' STR
SG ::= 'go' 'to' I
SI ::= 'if' A 'then' S
SP ::= 'print' STR
SM ::= 'move' ('left' | 'right') OS
SE ::= ''
SC ::= '{' L '}'
```

5. We finally obtain:
```
I  ::= [a-z]+
OS ::= 'one-square'
DOT::= '.'
```

```
LD  ::= [a-z]+ (':' LD)?
DL  ::= [a-z]+ (',' DL)?
STR ::= '"' I '"'
A   ::= 'the-tape-symbol' 'is' STR

SG  ::= 'go' 'to' I
SI  ::= 'if' A 'then' S
SP  ::= 'print' STR
SM  ::= 'move' ('left' | 'right') OS
SE  ::= ''
SC  ::= '{' L '}'
S   ::= (LD ':')? (SP | SM | SG | SE | SI  | SC)
L   ::= S(';' L)?
P   ::= 'tape-alphabet'  'is'  DL  ';'   L DOT
```

A completely named schema for Turingol corresponds to the rules provided.

**Appendix 2: Finite graphs. Properties, actions, instructions**

We will need some simple propositions for work with finite oriented labeled graphs.
For example, we will need the following proposition:
"There exists a unique 'W1' arrow", where W1 is any word. This proposition is true if one and only one arrow having a word W1 as its label exists on the entire graph.
In addition, we will need actions, including actions changing the graph.

*The normal execution condition* is graph property required for an action (proposition) to be executed (evaluated) normally. Some combination of normal execution conditions specifies the necessary and sufficient condition of normal execution (evaluation).
Note: Post describes situations where the Executor may be failed. True, Post speaks about inapplicability ("assuming" [6]). However, failure is an important element of programmer's thinking – they do occur, and he must think about fail-safety.
Post describes only two failure situations:
- Attempt to write in a cell already containing a value.
- Attempt to delete a value from a cell containing no such value.

Let's introduce a *path formula* to indicate a node on a graph:
Let WORDN mean the label assigned to the unique graph node (let's designate it N1), and WORD mean any word. Then
- WORDN+WORD means the node located at the end of the arrow starting from N1 and labeled by the word WORD,
- WORDN-WORD means the node in the origin of the arrow reaching N1 and labeled by the word WORD.

It means that we follow the arrow in the first case and move in the opposite direction in the second case. If there are several arrows for jumping, or there is no one, then the formula is considered inapplicable, and an attempt to use such formula during execution of the algorithm will result in a crash.
The initial node must be the only node on the graph. For example, if the path formula starts with a 'tape-alphabet' labeled node, then a crash is possible, as there exist more than one specified node or there is no one.

If we are in a node that is considered to be a current node, then +WORD means a node located at the end of the WORD arrow starting at the current node, and -WORD means a node located at the start of the WORD arrow ending at the current node.

If a word contains the '-' letter or is an empty word, then it has to be put in quotation marks, i.e. an empty word will be designated as '' in the path formula.

*Path passability*: Let P1 be a path formula. We will need an proposition of the type "P1 path is passable" from the current or a given node.

Some elementary propositions using path formulas are provided below. Necessary and sufficient conditions of normal execution are also provided for these propositions.

Let W1 designate a word, P1, P2 be some path formulas.

**Tab. 2** propositions. condition of normal execution

| proposition | Necessary and sufficient condition of normal execution |
|---|---|
| The P1 node label equals P2 node label. | P1 path is passable and P2 path is passable. |
| No 'W1' arrow exists to the P1 node. | P1 path is passable. |
| No 'W1' arrow exists from the P1 node. | P1 path is passable. |

*Actions*: Some actions required for work with oriented graphs and their necessary and sufficient conditions of normal execution are listed below.

Let W1 be a word, P1, P2 be some path formulas.

**Tab. 3** actions. conditions of normal execution

| Action phrase | Necessary and sufficient condition of normal execution | Description |
|---|---|---|
| label the P1 node by the P2 node label | P1 path is passable and P2 path is passable. | The label of the target node is overwritten. |
| reassign the 'W1' arrow to the P1 node | P1 path is passable and there exists a unique W1 arrow. | The arrow end is reassigned. |
| create a node and an arrow from it to the P2 node | P2 path is passable. | The arrow and node created are labeled by an empty word. |
| create a node and an arrow from P2 node to it | P2 path is passable. | The arrow and node created are labeled by an empty word. |

| Action phrase | Necessary and sufficient condition of normal execution | Description |
|---|---|---|
| follow the 'W1' arrow | There exists a unique W1 arrow from the current node. | Note: A crash is possible in two cases:<br>1. There exists no arrow with such label<br>2. There exist several arrows with the same label |

*Direction*: Any phrase of an action represented as a sentence is a direction.
Let S1 be a proposition, D1 - an action phrase. Then
"If S1, then D1." is a direction.

*Note*: According to Post, direction is a numbered sentence in English.
He lists three forms of direction [6], p.103-104: "Start at the starting point and follow direction 1. It is then to consist of a finite number of directions to be numbered 1, 2, 3... n. The i-th direction is then to have one of the following forms:
(A) Perform operation Oi [Oi = (a), (b), (c), or (d)] and then follow direction $j_i$,
(B) Perform operation (e) and according as the answer is yes or no correspondingly follow direction $j_i\,''$ or $j_i\,''$,
(C) stop."

*Note on the starting point*: According to Post, a one-dimensional space of boxes is infinite in both directions, and one of the box is selected as the initial one – this is the box where the worker begins to work.

*Instruction* is a sequence consisting of one or more directions.
Note: Post's directions (A) and (B) forms can be represented as follows:
(A) Perform operation $O_i$. Follow direction $j_i$.
(B) If (e) then follow direction $j_{i1}$. Follow direction $j_{i2}$.
Then they will consist of two simpler directions.
An instruction is obviously a special case of an algorithm.